\documentclass[12pt]{article}

\usepackage[dvips]{graphicx}
\usepackage{epsfig}
\usepackage{amsmath,amsfonts,amssymb,amsthm}
\usepackage{verbatim}
\usepackage{psfrag}
\usepackage{bm}
\usepackage{bbm}
\usepackage[square,comma,sort&compress,numbers]{natbib}
\usepackage{color}
\usepackage{slashed}

\usepackage{epsf,epsfig}
\usepackage{graphics}

\begin{document}
\thispagestyle{empty}

\begin{flushright}
{
\small
TTK-11-37\\
TUM-HEP 816/11\\
}
\end{flushright}

\vspace{0.4cm}
\begin{center}
\Large\bf\boldmath
Finite Width in out-of-Equilibrium Propagators and Kinetic Theory
\unboldmath
\end{center}

\vspace{0.4cm}

\begin{center}
{Bj\"orn~Garbrecht}\\
\vskip0.2cm
{\it Institut f\"ur Theoretische Teilchenphysik und Kosmologie,\\ 
RWTH Aachen University, 52056 Aachen, Germany}\\
\vskip.2cm
\end{center}
\begin{center}
and\\
\vskip.3cm
{Mathias~Garny}\\
\vskip0.3cm
{\it Physik-Department T30d, Technische Universit\"at M\"unchen,\\
James-Franck-Stra{\ss}e, 85748 Garching, Germany}\\
\vskip1.4cm
\end{center}

\begin{abstract}

We derive solutions to the Schwinger-Dyson equations
on the Closed-Time-Path for a scalar field in the limit where
backreaction is neglected. In Wigner space, the
two-point Wightman functions
have the curious property that the equilibrium component
has a finite width, while the out-of equilibrium component has
{\it zero} width. This feature is confirmed in a numerical simulation
for scalar field theory with quartic interactions. When substituting
these solutions into the collision term, we observe that an expansion
including terms of {\it all} orders in gradients leads to an
{\it effective} finite-width. Besides, we observe no
breakdown of perturbation theory, that is sometimes
associated with pinch singularities.
The effective width is identical
with the width of the equilibrium component. Therefore, reconciliation
between the {\it zero}-width behaviour and the usual notion
in kinetic theory, that the out-of-equilibrium contributions have
a finite width as well, is achieved. This result may also be viewed as
a generalisation of the fluctuation-dissipation relation to
out-of-equilibrium systems with negligible backreaction.

\end{abstract}

\newpage

\section{Introduction}

Observables in finite-density systems are typically derived
from expectation values of operators that are not time-ordered.
In equilibrium field theory,
these expectation values can be obtained from analytic
continuation of the corresponding expectation values calculated
in the Matsubara imaginary time formalism. Alternatively,
one may calculate the
expectation values directly in real time. For this purpose,
diagrammatic rules can be formulated that are known as the
Closed-Time-Path (CTP) formalism \cite{Schwinger:1960qe,Keldysh:1964ud}.
For out-of-equilibrium systems, the Matsubara formalism does
not apply and the CTP formalism is the standard method
for a description derived from first
principles of Quantum Field
Theory~\cite{Calzetta:1986cq,Chou:1984es,Berges:2004yj}.

While a canonical formulation is also possible~\cite{Maldacena:2002vr},
the CTP Feynman rules can easily be derived from a 
CTP generating functional~\cite{Calzetta:1986cq}. Moreover, a
particularly useful variant of the functional approach is to 
use the two-particle-irreducible (2PI) effective action
in order to derive
Schwinger-Dyson equations. These are self-consistent equations
for the dressed propagators $\Delta(x,y)$
and in principle encompass the full
real time evolution of the quantum system up to two-loop order
in the effective action~\cite{Calzetta:1999xh,Berges:2004pu}.

Analytical approaches typically base on linear 
approximations to the Schwinger-Dyson equations.
The linearisation may be realised by
neglecting the functional dependence of the
self-energy $\Pi(x,y)$ on the propagators $\Delta(x,y)$, which
implies the neglection of backreaction.
Moreover, it is often assumed that $\Pi(x,y)$ is invariant
under space-time translations, as it is appropriate for
a homogeneous, time-independent background.
The linearisation can therefore be justified by the gradient expansion
\cite{Greiner:1998vd,Prokopec:2003pj,Prokopec:2004ic} or by the notion
of a bath, that does not mediate a backreaction
to the out-of-equilibrium test field~\cite{Anisimov:2008dz,Anisimov:2010aq,Drewes:2010pf,Anisimov:2010dk}.
These approximations
are often applicable to weakly coupled systems
close-to-equilibrium~\cite{Calzetta:1986cq,CalzettaHuBook}.

One could therefore argue that instead of the
CTP-formalism,
one might as well use
a heuristic and more intuitive Boltzmann approach in order
to formulate kinetic equations that can be solved analytically,
or at least semi-analytically. Moreover, such Boltzmann equations
can straightforwardly be modified in order to account for a
finite width of the interacting quasi-particles and in order
to account for Fermi-Dirac or Bose-Einstein statistics.
For finite density QCD,
a powerful kinetic theory that includes off-shell effects
is formulated this way in Refs.~\cite{Arnold:2000dr,Arnold:2001ba,Arnold:2001ms,Arnold:2002ja,Arnold:2002zm,Arnold:2003zc}.

The present work is motivated by a number of recent applications
of the CTP formalism to leptogenesis and electroweak baryogenesis~\cite{Konstandin:2005cd,Cirigliano:2009yt,Anisimov:2010aq,Anisimov:2010dk,Garny:2009rv,Garny:2009qn,Garny:2010nj,Beneke:2010wd,Beneke:2010dz,Garny:2010nz,Garbrecht:2010sz,Cirigliano:2011di}.
In single-flavour leptogenesis, the CTP formulation naturally
removes the non-unitary
overcounting of real intermediate states, that
has to be performed explicitly when using the heuristic Boltzmann ansatz.
Moreover, the CTP-formalism readily provides a framework for the
description of flavour coherence. It can be successfully applied
to flavour leptogenesis~\cite{Beneke:2010dz} and to models
of electroweak baryogenesis
from flavour mixing~\cite{Konstandin:2005cd,Cirigliano:2009yt,Cirigliano:2011di}.
The approach bears the promise of further
progress in the directions of $N_2$-leptogenesis, leptogenesis in
the weak washout regime and resonant leptogenesis.

These applications show clear advantages over the conventional Boltzmann
approach and
justify a further development of the kinetic theory
based on the CTP formalism.
The improvements from the application of CTP methods
to leptogenesis
beyond the Boltzmann approach additionally
encompass thermal corrections in the
cut propagators of intermediate on-shell particles~\cite{Garny:2009rv,Garny:2009qn,Anisimov:2010aq,Garny:2010nj,Beneke:2010wd}
and the treatment of
flavour coherence~\cite{Beneke:2010dz}.
It is then natural to calculate corrections beyond these
leading order approximations. These can originate from higher order derivatives~\cite{Garny:2010nz} or from considering the effect of the finite width of the
singlet neutrino, that decays out of
equilibrium~\cite{Anisimov:2010aq,Anisimov:2010dk}.

In Refs.~\cite{Anisimov:2008dz,Anisimov:2010aq,Drewes:2010pf,Anisimov:2010dk}, the Schwinger-Dyson equations in the approximation
of negligible backreaction are solved in order to obtain the propagator
of an out-of-equilibrium field that includes finite-width corrections.
We emphasise that the assumption of a stationary bath is equivalent to any 
approximation scheme where the dependence of the self-energy
of the out-of-equilibrium field on the average time
and on 
derivatives with respect to average time are neglected,
such as some variants of the gradient expansion.
In the case of leptogenesis, the out-of-equilibrium field is a
heavy singlet neutrino and the background bath consists of
the $O(100)$ Standard Model degrees of freedom that are in
equilibrium during the time of leptogenesis.

Neglecting the backreaction leads to a somewhat curious form
of the Wightman functions for the out-of equilibrium field
[{\it cf.} Eqs.~(\ref{distr:singular},\ref{distr:singular:coordinate}) below]. (The Wightman functions are
the two-point correlations without time ordering. They
contain statistical information about the system,
{\it cf.} Appendix~\ref{appendix:twopoint} and
Refs.~\cite{Greiner:1998vd,Prokopec:2003pj}.)
First, there is
a contribution corresponding to a quasi-particle distribution that
is in equilibrium with the background bath. This distribution has
a finite width around the quasi-particle pole, which is given by the
relaxation rate of the system toward equilibrium, in accordance with the
fluctuation-dissipation theorem for equilibrium systems. In coordinate space,
the finite width can be identified with an exponential decay in the relative
coordinate of the Wightman function ${\rm i}\Delta^{<,>}(x,y)$.
Second, there is an out-of equilibrium
contribution that exhibits no decay in the relative coordinate, or equivalently
{\it zero} width.

There is an interesting connection of the solutions
[Eq.~(\ref{distr:singular}) below], that decompose into a
finite-width equilibrium and a zero width out-of equilibrium
contribution, with the problem of pinch singularities
in out-of-equilibrium field theory in the real time formalism.
In Fourier space, it turns out that all self-energy insertions
resum in such a way that the equilibrium contribution attains a finite width,
while the out-of-equilibrium terms cancel~\cite{Altherr:1994jc}. We briefly review this in Section~\ref{section:pinch}. It is sometimes argued
that the occurrence of pinch singularities signals a breakdown
of perturbation theory for out-of-equilibrium systems, even when
the coupling constant is arbitrarily small~\cite{Greiner:1998ri,Boyanovsky:1999cy}. We emphasise however,
that within the Schwinger-Dyson approach, no such problem occurs.

The zero-width behaviour also appears to be related
to a feature of the Schwinger-Dyson equations on the CTP
that is discussed
in Refs.~\cite{Greiner:1998vd,Calzetta:1999xh}.
When interpreting
the kinetic equations~(\ref{kinetic:allorders}),
that appear as a subset of the Schwinger-Dyson equations,
as Langevin equations, they obviously feature a Stokes term, that induces the
approach of the macroscopic observables to equilibrium, but they lack a
Langevin term, that induces fluctuations and might
lead to a finite width for
the non-equilibrium distribution. In Ref.~\cite{Greiner:1998vd}, the kinetic equations
are reformulated in a way such that they take a form similar to Langevin 
equations. Since the resulting equations are however equivalent to the original
kinetic equations in the stationary bath approximation, the feature of
{\it zero} width distributions still persists within the solutions.
The truncation of $n$-point functions is proposed as the reason for the
absence of a Langevin term in Ref.~\cite{Calzetta:1999xh}.
Clearly, when the effects of backreaction are neglected,
the functional dependence
of the self-energies on the two-point functions is dropped
in order to achieve a linearisation of the Schwinger-Dyson equations.
However, while stochastic sources for the fluctuations of the propagators
$\sim\langle\Delta \Delta\rangle -\Delta \Delta$ are addressed,
the zero width contributions to $\Delta^{<,>}$
are not discussed in Ref.~\cite{Calzetta:1999xh}.

The occurrence of zero-width terms
appears counter-intuitive. Based on the expectation that
one should not be able to tell whether a quasi-particle is part of the
equilibrium or non-equilibrium distribution by measuring its width,
rather than the solution~(\ref{distr:singular}), one might use the
ansatz~(\ref{distr:smooth}), where all quasi-particles have a
finite width. Na\"ively, we might
expect important consequences for calculations, that rely on the heuristic
ansatz~(\ref{distr:smooth}),
where both the equilibrium and the non-equilibrium contributions to the quasi-particle distributions
have a finite width, as {\it e.g.} in Refs.~\cite{Arnold:2000dr,Arnold:2001ba,Arnold:2001ms,Arnold:2002ja,Arnold:2002zm,Arnold:2003zc,Mrowczynski:1997hy,Garbrecht:2008cb}, where off-shell
contributions to the scattering rates are of leading importance.

A purpose of the present work is to resolve this seeming contradiction.
In Section~\ref{section:pinch}, we review the summation of pinch
singularities in the CTP formalism. The propagator with zero-width
out-of-equilibrium contributions in the limit of negligible backreaction
is derived in Section~\ref{section:KB}. An explicit numerical check of
these solutions for a scalar theory with quartic interactions and a
single excited mode is performed in Section~\ref{section:num}.
The Figures in that Section
illustrate the behaviour of the non-equilibrium and equilibrium
components of the statistical propagator (the sum of the
two Wightman functions) in two-time representation.
These discussions confirm the validity of the solution~(\ref{distr:singular})
for the Wightman function with a zero-width out of equilibrium component
in situations with negligible backreaction. The agreement between
numerical and analytical results is of great value, because we
can exclude that the zero-width behaviour is due to analytical
approximations that are not understood.
In Section~\ref{section:effectivewidth}, we show that the summation
of all gradient terms, that act on the on-shell $\delta$-function
leads to an effective finite-width behaviour of the out-of-equilibrium
distributions in the collision term. In order to ``observe'' the
finite width, in Section~\ref{section:thresholds}, we propose to use
interactions that are kinematically forbidden for zero-width particles.
The main conclusion of Sections~\ref{section:effectivewidth}
and~\ref{section:thresholds} is that the use of the
solution~(\ref{distr:singular}) with the zero-width components and the
summation of the gradient terms is equivalent to the use of the
heuristic ansatz~(\ref{distr:smooth}) without summation of the gradient
terms. This way, the solution~(\ref{distr:singular}) is reconciled
with heuristic approaches, that assume a finite width for the out-of-equilibrium
distributions. We may also interpret this result as a generalisation
of the fluctuation-dissipation relation to out-of-equilibrium systems in
situations, when backreaction is negligible. Further discussions and
conclusions are offered in Section~\ref{section:conclusions}.

\section{Pinch Singularities}
\label{section:pinch}

The analysis of this Section is originally reported in
Ref.~\cite{Altherr:1994jc}. Here, we recapitulate it
in order to put it more directly within the
present context. We consider a real scalar field
$\varphi$ of mass $m_\varphi$ within the CTP formalism. In this Section,
we evaluate the propagator, which is a two-point function,
in Fourier (momentum) space. As an
implicit consequence, we do not allow for a dependence of the
propagator on the average coordinate, {\it i.e.} we
cannot expect it to describe the approach of the field $\varphi$
toward equilibrium.
The tree-level
CTP propagator for the scalar field can then
be expressed in matrix form as
\begin{align}
\label{Delta:tree}
{\rm i}\Delta_\varphi^{(0)}(k)&=
\left(
\begin{array}{cc}
{\rm i}\Delta_\varphi^{(0)++}(k) & {\rm i}\Delta_\varphi^{(0)+-}(k)\\
{\rm i}\Delta_\varphi^{(0)-+}(k) & {\rm i}\Delta_\varphi^{(0)--}(k)
\end{array}
\right)
\\\nonumber
&={\rm i}\Delta_\varphi^{(0)R}(k)
\left(
\begin{array}{cc}
1+f_\varphi(k) & f_\varphi(k)\\
1+f_\varphi(k) & f_\varphi(k)
\end{array}
\right)
-{\rm i}\Delta_\varphi^{(0)A}(k)
\left(
\begin{array}{cc}
f_\varphi(k) & f_\varphi(k)\\
1+f_\varphi(k) & 1+f_\varphi(k)
\end{array}
\right)\,,
\end{align}
where
\begin{align}
f_\varphi(k^0,\mathbf k)\equiv f_\varphi(k)=\left\{
\begin{array}{ll}
f_\varphi(\mathbf k) & \textnormal{for}\quad k^0>0\\
-1-f_\varphi(-\mathbf k) & \textnormal{for}\quad k^0<0
\end{array}
\right.
\end{align}
and
$f_\varphi(\mathbf k)$ is the distribution function in momentum space,
and
\begin{align}
{\rm i}\Delta_\varphi^{(0)R,A}(k)=\frac{\rm i}{k^2-m^2_\varphi(k)\pm{\rm i}k^0\varepsilon}
\end{align}
are the tree-level retarded and advanced propagators, with
$\varepsilon$ being infinitesimal.
For the definition of the various two-point functions on the CTP
and their relations amongst one another, {\it cf.}
Appendix~\ref{appendix:twopoint}.
The tree-level Wightman functions have {\it zero} width:
\begin{subequations}
\begin{align}
{\rm i}\Delta_\varphi^{(0)<}(k)={\rm i}\Delta_\varphi^{(0)+-}(k)&=
2\pi \delta(k^2-m_\varphi^2){\rm sign}(k^0)f_\varphi(k)
\\\notag
&=2\pi \delta(k^2-m_\varphi^2)\left[
\vartheta(k_0) f_\varphi(\mathbf k)
+\vartheta(-k_0) (1+f_\varphi(-\mathbf k))\right]
\,,
\\
{\rm i}\Delta_\varphi^{(0)>}(k)={\rm i}\Delta_\varphi^{(0)-+}(k)&=
2\pi \delta(k^2-m_\varphi^2){\rm sign}(k^0)[1+f_\varphi(k)]
\\\notag
&=2\pi \delta(k^2-m_\varphi^2)\left[
\vartheta(k_0) (1+f_\varphi(\mathbf k))
+\vartheta(-k_0) f_\varphi(-\mathbf k)\right]
\,.
\end{align}
\end{subequations}
We note that the zero-width spectral function
\begin{align}
2\Delta_\varphi^{(0)\cal A}(k)=2\pi \delta(k^2-m_\varphi^2){\rm sign}(k^0)
\end{align}
occurs as a factor within these expressions.

We aim to obtain Wightman functions of finite width by summing over all
insertions of self energies ${\rm i}\Pi_\varphi$, which are understood to be CTP
matrices. For this purpose,
we define the matrix
\begin{align}
\Xi=
\left(
\begin{array}{cc}
1 & 0\\
0 & -1
\end{array}
\right)\,,
\end{align}
to take account account of factors $\pm 1$ that occur in the CTP
Feynman rules. The self-energies ${\rm i}\Pi_\varphi$
on the CTP are defined in analogy with
Eq.~(\ref{Delta:tree}) . For $N$ insertions in the tree
propagator, we obtain
\begin{align}
{\rm i}\Delta_\varphi^{(0)} &
\left(
\Xi {\rm i}\Pi_\varphi \Xi {\rm i}\Delta_\varphi^{(0)}
\right)^N
\\\nonumber
&=
{\rm i}\Delta_\varphi^{(0)R} \left({\rm i}\Pi_\varphi^R {\rm i}\Delta_\varphi^{(0)R}\right)^N
\left(
\begin{array}{cc}
1+f_\varphi & f_\varphi\\
1+f_\varphi & f_\varphi
\end{array}
\right)
-{\rm i}\Delta_\varphi^{(0)A} \left({\rm i}\Pi_\varphi^A {\rm i}\Delta_\varphi^{(0)A}\right)^N
\left(
\begin{array}{cc}
f_\varphi & f_\varphi\\
1+f_\varphi & 1+f_\varphi
\end{array}
\right)
\\\nonumber
&-
{\rm i}\Delta_\varphi^{(0)R}{\rm i}\Delta_\varphi^{(0)A}
\left[
f_\varphi {\rm i}\Pi_\varphi^>-(1+f_\varphi){\rm i}\Pi_\varphi^<
\right]
\sum\limits_{j=0}^{N-1}({\rm i}\Delta_\varphi^{(0)R} {\rm i}\Pi_\varphi^R)^{j}
({\rm i}\Delta_\varphi^{(0)A}{\rm i}\Pi_\varphi^A)^{N-1-j}
\left(
\begin{array}{cc}
1 & 1\\
1 & 1
\end{array}
\right)
\,,
\end{align}
where we have suppressed the arguments $k$.
The last term contains the so-called pinch singularities, since
products of poles with the same real parts but opposite
imaginary parts occur there.
From this expression, we obtain the resummed, full propagator
\begin{align}
{\rm i}\Delta_\varphi&=
\sum\limits_{N=0}^\infty
{\rm i}\Delta_\varphi^{(0)}
\left(
\Xi {\rm i}\Pi_\varphi \Xi {\rm i}\Delta_\varphi^{(0)}
\right)^N
\\\nonumber
&=
\frac{\rm i}{k^2-m_\varphi^2-\Pi_\varphi^R}\left(
\begin{array}{cc}
1+f_\varphi & f_\varphi\\
1+f_\varphi & f_\varphi
\end{array}
\right)
-
\frac{\rm i}{k^2-m_\varphi^2-\Pi_\varphi^A}\left(
\begin{array}{cc}
f_\varphi & f_\varphi\\
1+f_\varphi & 1+f_\varphi
\end{array}
\right)
\\\nonumber
&
-
\frac{\rm i}{k^2-m_\varphi^2-\Pi_\varphi^R}\frac{\rm i}{k^2-m_\varphi^2-\Pi_\varphi^A}
\left[
f_\varphi {\rm i}\Pi_\varphi^>-(1+f_\varphi){\rm i}\Pi_\varphi^<
\right]
\left(
\begin{array}{cc}
1 & 1\\
1 & 1\\
\end{array}
\right)
\,,
\end{align}
where the last term results from the resummation of the pinch singularities.
When we make use of $\Pi_\varphi^{\cal A}=\frac{\rm i}2\left(\Pi_\varphi^>-\Pi_\varphi^<\right)=\frac{\rm i}{2}\left(\Pi_\varphi^R-\Pi_\varphi^A\right)$
({\it cf.} Appendix~\ref{appendix:twopoint})
and define the stationary distribution
function
\begin{align}
f_\varphi^{\rm st}(k)=\frac{\Pi_\varphi^<(k)}{\Pi_\varphi^>(k)-\Pi_\varphi^<(k)}
\,,
\end{align}
the resummed propagator is given by
\begin{align}
\label{Delta:equilibrium}
{\rm i}\Delta_\varphi(k)
=&
\frac{\rm i}{k^2-m_\varphi^2-\Pi_\varphi^R}
\left(
\begin{array}{cc}
1+f_\varphi^{\rm st}(k) & f_\varphi^{\rm st}(k)\\
1+f_\varphi^{\rm st}(k) & f_\varphi^{\rm st}(k)
\end{array}
\right)
\\\nonumber
-&\frac{\rm i}{k^2-m_\varphi^2-\Pi_\varphi^A}
\left(
\begin{array}{cc}
f_\varphi^{\rm st}(k) & f_\varphi^{\rm st}(k)\\
1+f_\varphi^{\rm st}(k) & 1+f_\varphi^{\rm st}(k)
\end{array}
\right)\,.
\end{align}
In particular, the Wightman functions are now
\begin{subequations}
\begin{align}
{\rm i}\Delta_\varphi^{<}(k)&={\rm i}\Delta_\varphi^{+-}(k)=
\frac{2\Pi_\varphi^{\cal A}}{(k^2-m_\varphi^2-\Pi_\varphi^H)^2+{\Pi_\varphi^{\cal A}}^2}
f_\varphi^{\rm st}(k)
=2\Delta_\varphi^{\cal A}(k) f_\varphi^{\rm st}(k)
\,,
\\
{\rm i}\Delta_\varphi^{>}(k)&={\rm i}\Delta_\varphi^{+-}(k)=
\frac{2\Pi_\varphi^{\cal A}}{(k^2-m_\varphi^2-\Pi_\varphi^H)^2+{\Pi_\varphi^{\cal A}}^2}
[1+f_\varphi^{\rm st}(k)]
=2\Delta_\varphi^{\cal A}(k)[1+f_\varphi^{\rm st}(k)]
\,,
\end{align}
\end{subequations}
which have the desired finite width for $\Pi_\varphi^{\cal A}\not=0$.
But evidently, there
is the unwanted feature that
all dependence on the distribution function $f_\varphi(k)$ has vanished
in a spectacular way.
Instead, the propagator depends on $f_\varphi^{\rm st}(k)$, which is determined by the self-energy $\Pi_\varphi(k)$.
This is a consequence of the fact that the calculation has been performed
in Fourier space, with no provision for a time dependence that would describe the
relaxation of the system toward equilibrium. The only possible solution
is the one that is in equilibrium with the ``background'' described by
the self-energy $\Pi_\varphi$.

The Schwinger-Dyson equations in Wigner
space in principle account for the time dependence.
In Section~\ref{section:effectivewidth},
we show that in order to capture the finite-width behaviour for
the time-dependent out-of-equilibrium contributions, one must
include terms of {\it all} orders in the gradient expansion as well.
The finite width of the out-of-equilibrium distribution appears
not as a feature of the propagator itself, but it is rather
encompassed in the collision term expanded to all orders in the
gradient expansion.

\section{Solutions to the Kadanoff-Baym Equations without Backreaction}
\label{section:KB}

In this Section, we solve the Schwinger Dyson Equations
on the CTP under the assumption that all self-energies $\Pi_\varphi(x,y)$
are
time-translation invariant. (In Wigner space, this implies that
they are time-independent).
This amounts to an approximation
where the backreaction of the field $\varphi$ on the
fields that are relevant for the
self-energies $\Pi_\varphi$ is neglected.
We derive the solutions in Wigner space and thereby recover the
result that has been found earlier in two-time representation
within Ref.~\cite{Anisimov:2008dz}.

For a real scalar field $\varphi$ of constant mass $m_\varphi$,
the Kadanoff-Baym equations in Wigner space are
\begin{align}
\label{KB:allorders}
&\left[
k^2-\frac14 \partial^2_t +{\rm i} k^0\partial_t -m_\varphi^2
\right]\Delta_\varphi^{<,>}
-{\rm e}^{-{\rm i}\diamond}\{\Pi_\varphi^H\}\{\Delta_\varphi^{<,>}\}
-{\rm e}^{-{\rm i}\diamond}\{\Pi_\varphi^{<,>}\}\{\Delta_\varphi^{H}\}
\\\notag
&\hskip1cm =\frac12{\rm e}^{-{\rm i}\diamond}
\left(
\{\Pi_\varphi^>\}\{\Delta_\varphi^<\}
-\{\Pi_\varphi^<\}\{\Delta_\varphi^>\}
\right)
\end{align}
and the equations for the retarded and advanced propagators
\begin{align}
\label{retav:Wigner}
\left[
k^2+{\rm i}k^0\partial_t-\frac14 \partial_t^2 -m_\varphi^2
\right]\Delta_\varphi^{R,A}
-{\rm e}^{{\rm i}\diamond}\{\Pi_\varphi^H\}\{\Delta_\varphi^{R,A}\}
\pm{\rm e}^{-{\rm i}\diamond}\{{\rm i}\Pi_\varphi^{\cal A}\}
\Delta^{R,A}=1\,,
\end{align}
where the diamond operator is defined as
\begin{align}
\diamond\{A\}\{B\}=
\frac12\left(\partial_x A\right)\left(\partial_k B\right)
-\frac12\left(\partial_k A\right)\left(\partial_x B\right)\,,
\end{align}
and where $t=x^0$.
Provided the scale of macroscopic variations $\sim\partial_x$
is smaller than the typical energy-scale $\sim(\partial_k)^{-1}$,
we may expect that we can truncate at a finite order in the diamond
operator. This approximation scheme is known as the gradient expansion.
We refer to the right-hand side of Eq.~(\ref{KB:allorders}) as the collision
term. In Section~\ref{section:effectivewidth}, we note that the identification
of $\sim(\partial_k)^{-1}$ with the energy scale does not hold, when the
derivative acts on an on-shell $\delta$-function, but also that these terms can be 
straightforwardly summed and interpreted.

It is useful to split the Kadanoff-Baym equations into
constraint and kinetic equations as
\begin{subequations}
\label{Schwinger:Dyson:allorders}
\begin{align}
\label{constraint:allorders}
&\left[
k^2-\frac14 \partial^2_t  -m_\varphi^2
\right]\Delta_\varphi^{<,>}
-\cos(\diamond)\{\Pi_\varphi^H\}\{\Delta_\varphi^{<,>}\}
-\cos(\diamond)\{\Pi_\varphi^{<,>}\}\{\Delta_\varphi^{H}\}
\\\notag
&\hskip1cm =-\frac{\rm i}2\sin(\diamond)
\left(
\{\Pi_\varphi^>\}\{\Delta_\varphi^<\}
-\{\Pi_\varphi^<\}\{\Delta_\varphi^>\}
\right)\,,
\\
\label{kinetic:allorders}
& k^0\partial_t
 {\rm i}\Delta_\varphi^{<,>}
+\sin(\diamond)\{\Pi_\varphi^H\}\{\Delta_\varphi^{<,>}\}
+\sin(\diamond)\{\Pi_\varphi^{<,>}\}\{\Delta_\varphi^{H}\}
\\\notag
&\hskip1cm =-\frac12\cos(\diamond)
\left(
\{{\rm i}\Pi_\varphi^>\}\{{\rm i}\Delta_\varphi^<\}
-\{{\rm i}\Pi_\varphi^<\}\{{\rm i}\Delta_\varphi^>\}
\right)\,.
\end{align}
\end{subequations}

The assumption of a constant $\Pi_\varphi$ amounts to neglecting
its functional dependence on $\Delta_\varphi$, and
Eqs.~(\ref{KB:allorders},\ref{retav:Wigner})
become linear, such that they can be solved using straightforward
methods. Of course, this linearisation corresponds to the neglection
of backreaction.
In particular, the
result for the stationary propagator
from the resummation of pinch singularities is reproduced easily:
Eq.~(\ref{retav:Wigner}) yields
\begin{subequations}
\begin{align}
\label{retav}
\Delta_\varphi^{R,A}&=\frac1
{k^2-m_\varphi^2-\Pi_\varphi^H\pm{\rm i}\Pi_\varphi^{\cal A}}\,,
\\
\Delta_\varphi^H&=\frac12(\Delta_\varphi^A+\Delta_\varphi^R)
\label{Delta:H}
=\frac{k^2-m_\varphi^2-\Pi_\varphi^H}{(k^2-m_\varphi^2-\Pi_\varphi^H)^2+{\Pi_\varphi^{\cal A}}^2}
\,,
\end{align}
\end{subequations}
which can be substituted in Eq.~(\ref{constraint:allorders}) such that
\begin{align}
\label{Delta:st}
{\rm i}\Delta_\varphi^{{\rm st}<,>}
&=\frac{
\Pi_\varphi^{\cal A}}{(k^2-m_\varphi^2-\Pi_\varphi^H)^2+{\Pi_\varphi^{\cal A}}^2}
\frac{{\rm i}\Pi_\varphi^{<,>}}{\frac{{\rm i}}{2}(\Pi_\varphi^>-\Pi_\varphi^<)}
=2\frac{
\Pi_\varphi^{\cal A}}{(k^2-m_\varphi^2-\Pi_\varphi^H)^2+{\Pi_\varphi^{\cal A}}^2}
f_\varphi^{{\rm st}<,>}
\,,
\end{align}
where we define
$f_\varphi^{{\rm st}<}=f_\varphi^{\rm st}$ and $f_\varphi^{{\rm st}>}=1+f_\varphi^{\rm st}$.
The resummation of pinch singularities
is implicit within the Schwinger-Dyson equations on the
Closed Time Path, such that their solution
is perhaps a simpler and more direct procedure for obtaining approximations
to the full propagators.


Under these assumptions, the Wigner space solution
to Eqs.~(\ref{KB:allorders}) and~(\ref{retav:Wigner}) at zeroth order
in the operator $\diamond$ is
\begin{align}
\label{distr:singular}
{\rm i}\Delta_\varphi^{<,>}(k,t)=&{\rm i}\Delta_\varphi^{{\rm st}<,>}(k)+
\delta f_\varphi(k^0,\mathbf{k}){\rm sign}(k^0)2\pi\delta(k^2-m_\varphi^2-\Pi_\varphi^H){\rm e}^{-(\Pi_\varphi^{\cal A}/k^0)t}
\\\notag
=&
{\rm i}\Delta_\varphi^{{\rm st}<,>}(k)+
\delta f_\varphi(\mathbf{k})2\pi\delta(k^2-m_\varphi^2-\Pi_\varphi^H){\rm e}^{-\Gamma_\varphi(\mathbf k)t}
\,,
\end{align}
where $\delta f_\varphi(k^0,\mathbf{k})$ is an integration constant.
Note that the solution is always decaying for both particle ($k^0>0$)
and antiparticles ($k^0<0$), because $\Pi_\varphi^{\cal A}(k^0,k)$
is odd in $k^0$.
The obvious interpretation of $\delta f_\varphi(\mathbf{k})$
is the deviation of the particle
distribution function from its equilibrium form.

For the second equality, we have used the Breit-Wigner approximation.
This corresponds to the assumption that it is sufficient to
know the value of $\Pi_\varphi^{\cal A}(k)$
at the quasi-particle pole, where
\[
k^0=\pm\omega_\varphi(\mathbf k)=\pm\sqrt{\mathbf k^2+m_\varphi^2+\Pi_\varphi^H(\omega_\varphi(\mathbf k),k)}\,.
\]
The approximation is valid when
only contributions close to the pole are important and $\Pi_\varphi^{\cal A}(k)$
is a smooth function in $k^0$.
In other words, we may approximate
\begin{align}
\label{Pi:Breit:Wigner}
\Pi_\varphi^{\cal A}(\Delta t,\mathbf k)=\frac 1{2\rm i} \omega_\varphi^2(\mathbf k)\Gamma_\varphi(\mathbf k)\,{\rm sign}(\Delta t)\,,
\end{align}
where $\omega_\varphi(\mathbf k)=\sqrt{\mathbf{k}^2+m_\varphi^2}$.
Then,
\begin{align}
\Pi_\varphi^{\cal A}(k^0,\mathbf k)=\frac{\omega_\varphi^2(\mathbf k)}{k^0}\Gamma_\varphi(\mathbf k)
\end{align}
is approximated by a function which is smooth in $k^0$
close to the quasi-particle pole. Here, $\Gamma_\varphi(\mathbf k)$
is the relaxation rate. In the present context, this approximation is
obviously valid, because the $\delta$-function forces the evaluation
of $\Pi^{\cal A}(k)$ at the quasi-particle poles.

The fact that $\Pi_\varphi^{\cal A}$, which appears within Eq.~(\ref{distr:singular})
as the finite width for the stationary (equilibrium)
distribution $f_\varphi^{\rm st}$,
and therefore describes fluctuations,
is the same as in the retarded propagator~(\ref{retav}), which describes
the dissipation of a linear perturbation toward equilibrium, is a consequence
of the celebrated fluctuation-dissipation relation for equilibrium
systems. In this light, the absence of fluctuations in the terms that
multiply $\delta f_\varphi$ in Eq.~(\ref{distr:singular}) may appear
as a surprise and suggest that the fluctuation-dissipation relation
cannot at all be generalised to out-of-equilibrium situations.
However,
in Section~\ref{section:effectivewidth}, we show that even though
the non-equilibrium contribution to ${\rm i}\Delta^{<,>}_\varphi$
in Eq.~(\ref{distr:singular}) does not exhibit a width, fluctuations
arise effectively from a summation of the gradients within the collision term.

The inverse Wigner transform of Eq.~(\ref{distr:singular})
then leads to the result
of Refs.~\cite{Anisimov:2008dz,Drewes:2010pf}:\footnote{
The solutions in Refs.~\cite{Anisimov:2008dz,Drewes:2010pf} also
encompass possible oscillatory terms in the average coordinate $t$
with the frequency $2\omega$. These can be identified with the
coherence solutions that are reported in
Refs.~\cite{Herranen:2008di,Herranen:2008hu,Herranen:2008hi}.
}
\begin{align}
\label{distr:singular:coordinate}
{\rm i}\Delta_\varphi^{<,>}(t;\Delta t)&\approx
\frac1{\omega_\varphi}{\rm e}^{-\frac{\Gamma_\varphi(\mathbf k){\rm sign}(\Delta t)}{2}\Delta t}
\left[
\cos(\omega_\varphi \Delta t) f_\varphi^{\rm st}(\mathbf{k}) +\frac 12{\rm e}^{\pm{\rm i}\omega_\varphi\Delta t}
\right]
\\\notag
&+\frac1{\omega_\varphi} \cos(\omega_\varphi \Delta t) {\rm e}^{-\Gamma_\varphi(\mathbf k) t}
\delta f(\mathbf{k})\,,
\end{align}
where we have
expanded to linear order in $\Pi_\varphi^{\cal A}$ within the exponentials
and the prefactors.
Apparently, the distribution~(\ref{distr:singular}) consists of a
finite-width equilibrium term and
a zero-width non-equilibrium contribution.
It na\"ively appears that we could tell whether a particle
belongs to the equilibrium part $f^{\rm st}$ or the non-equilibrium
part $\delta f$ by observing its width.
In the coordinate
representation~(\ref{distr:singular:coordinate}), this is
reflected by the fact that the correlations proportional to
the out-of equilibrium distribution $\delta f$ exhibit no
damping in the relative time coordinate $\Delta t$. A corresponding
behaviour is exhibited by the propagators for out-of-equilibrium
fermions reported in Refs.~\cite{Anisimov:2010aq,Anisimov:2010dk}, that
are derived under the same assumptions.

The solution~(\ref{distr:singular}) should be contrasted
with the ``Kadanoff-Baym ansatz'',
\begin{subequations}
\label{distr:smooth}
\begin{align}
{\rm i}\Delta_\varphi^{<}(k)&=
\frac{2\Pi_\varphi^{\cal A}}{(k^2-m_\varphi^2-\Pi_\varphi^H)^2+{\Pi_\varphi^{\cal A}}^2}
f_\varphi(k)
=2\Delta_\varphi^{\cal A}(k) f_\varphi(k)
\,,
\\
{\rm i}\Delta_\varphi^{>}(k)&=
\frac{2\Pi_\varphi^{\cal A}}{(k^2-m_\varphi^2-\Pi_\varphi^H)^2+{\Pi_\varphi^{\cal A}}^2}
[1+f_\varphi(k)]
=2\Delta_\varphi^{\cal A}(k)[1+f_\varphi(k)]
\,.
\end{align}
\end{subequations}
This relies on the intuitive expectation that it should not
be possible to distinguish between an equilibrium and a non-equilibrium
particle by measuring its width. In Section~\ref{section:thresholds},
we describe how such a measurement may be performed.
This ansatz is proposed more or less explicitly within the literature
on Kadanoff-Baym equations, see {\it e.g.} Eq.~(4.44) in Ref.~\cite{Greiner:1998vd},
Eq.~(11.13) in Ref.~\cite{CalzettaHuBook} or
Refs.~\cite{Garbrecht:2008cb,Berges:2004pu}.
In principle, it also underlies calculations
in more heuristic approaches to kinetic theory, where finite-width effects
are important, such as in Refs.~\cite{Arnold:2000dr,Arnold:2001ba,Arnold:2001ms,Arnold:2002ja,Arnold:2002zm,Arnold:2003zc}.
In the subsequent Section~\ref{section:num}, we numerically confirm
the correctness of the solution~(\ref{distr:singular}).
Nonetheless, as we show in Section~\ref{section:effectivewidth},
it turns out that the ansatz~(\ref{distr:smooth})
effectively resums an infinite number of gradients within the
collision term, such that its use in kinetic theory can be justified.

\begin{figure}
  \includegraphics{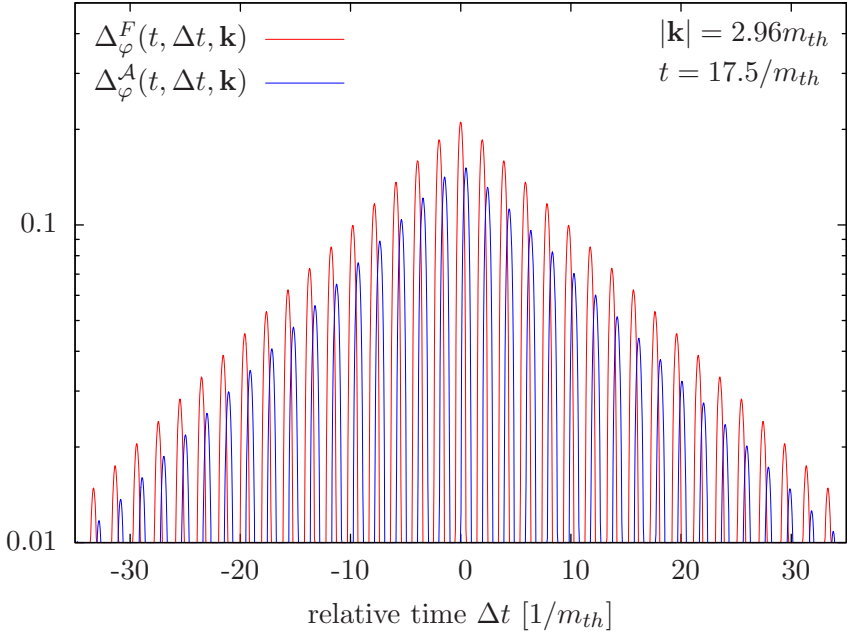}
  \caption{\label{fig2}  Statistical propagator $\Delta_\varphi^F=(\Delta_\varphi^>+\Delta_\varphi^<)/2$
  (red) and spectral function $\Delta_\varphi^{\cal A}=\frac{\rm i}2(\Delta_\varphi^>-\Delta_\varphi^<)$ (blue) obtained from a numerical
  solution of the Kadanoff-Baym equations. Shown is the dependence on the relative time $\Delta t$ for fixed
  central time $t$.}
\end{figure}

\begin{figure}
 \includegraphics{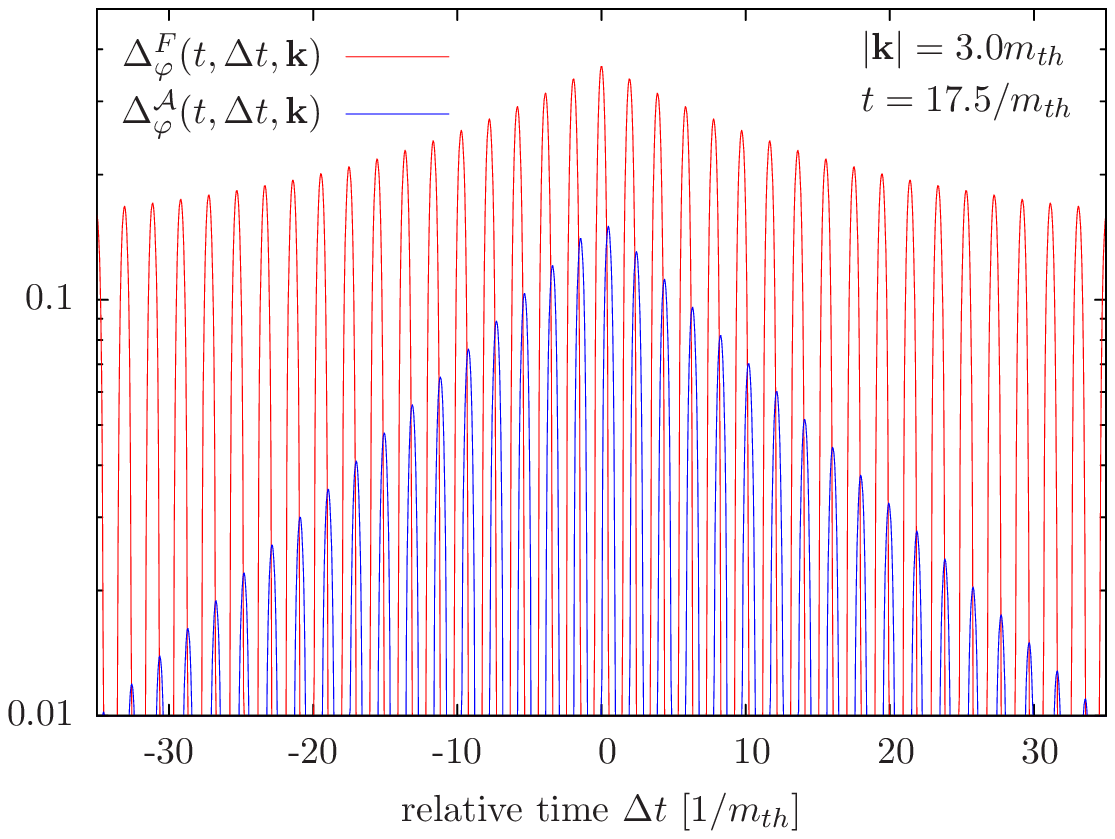}
 \caption{\label{fig1} As figure \ref{fig2}, but for the excited momentum mode $|\mathbf k|=3.0m_{th}$. The statistical
  propagator $\Delta_\varphi^F=(\Delta_\varphi^>+\Delta_\varphi^<)/2$ can be described by the sum of an exponentially
  damped equilibrium contribution and an undamped non-equilibrium contribution.}
\end{figure}

\section{Out-of-Equilibrium Propagator from a
Numerical Simulation}
\label{section:num}

In this Section, we present a numerical check that indeed,
when backreaction effects may be neglected,
Eq.~(\ref{distr:singular}) is a correct solution
to the Kadanoff-Baym equations,
while the na\"ive finite-width ansatz~(\ref{distr:smooth}) is
not.

The assumption that is crucial for deriving the analytical
solution~(\ref{distr:singular},\ref{distr:singular:coordinate}) to the Kadanoff-Baym equations
is the time-independence of the self-energy in Wigner space, $\Pi_\varphi(t,k_\mu)=\Pi_\varphi(k_\mu)$. This situation is realized
for example when the non-equilibrium field $\varphi$ interacts with a thermal bath. In this case, a
necessary condition for the time-independence of the self-energies is that the thermal bath can be
considered as an infinitely large reservoir, that is not influenced itself by the interaction with
the non-equilibrium field $\varphi$.

In this context the question arises how sensitive the analytical
solution~(\ref{distr:singular},\ref{distr:singular:coordinate}), and in particular
the peculiar behaviour of the non-equilibrium contribution, is to the underlying thermal-bath assumption.
In order to investigate this question, we compare the analytical expression with numerical solutions
to the complete Kadanoff-Baym equations, for which the self-energies are computed self-consistently.
This ensures that the full back-reaction is taken into account. Concretely,
we consider a real scalar field with $\lambda\varphi^4$ interaction, and solve the Kadanoff-Baym equations
in the two-time representation,
\begin{eqnarray}
 \left(\partial_x^2+m_\varphi^2+\frac{\lambda}{2}\Delta_\varphi^{<,>}(x,x)\right)\Delta_\varphi^{<,>
}(x,y) & = & 2\int_{t_0}^{y^0}d^4z\,\Pi_\varphi^{<,>}(x,z)\Delta_\varphi^{\cal
A}(z,y)\nonumber\\
 && {} - 2\int_{t_0}^{x^0}d^4z\,\Pi_\varphi^{\cal A}(x,z)\Delta_\varphi^{<,>}(z,y)  \;,
\nonumber\\
\end{eqnarray}
where $t_0$ is the initial time, $\Delta_\varphi^{\cal A}=\frac{\rm i}2(\Delta_\varphi^{>}-\Delta_\varphi^{<})$ is the spectral function, and
\[
  \int_{t_0}^{x^0}d^4z = \int_{t_0}^{x^0}dz^0 \int d^3z \;.
\]
For the self-energies, we use the so-called setting-sun approximation,
\begin{equation}
  \Pi_\varphi^{<,>}(x,y)= -\frac{\lambda^2}{6}\Delta_\varphi^{<,>}(x,y)^3 \;,
\end{equation}
which describes scattering processes, and is accurate for small enough coupling and occupation numbers
(see \cite{Lindner:2005kv,Berges:2000ur} for details). We assume that the
system is spatially homogeneous and isotropic. In this case it is convenient to transform to spatial momentum
space,
\begin{equation}
  \Delta_\varphi^{<,>}(t,\Delta t,|\mathbf{k}|) =  \int d^3x \, e^{-{\rm i}\mathbf{k}\cdot(\mathbf{x}-\mathbf{y})}\, \Delta_\varphi^{<,>}(x,y) \;,
\end{equation}
where, in addition, we have introduced the central time $t=(x^0+y^0)/2$ and the relative time $\Delta t=x^0-y^0$ for
convenience. The Wigner space functions would result by an additional Fourier transformation with
respect to $\Delta t$.

Since we are interested in situations where a non-equilibrium degree of freedom interacts with many other
degrees of freedom that are close to thermal equilibrium, we initialize the system at time $t=\Delta t=0$
with thermal correlators at temperature $T=1.5m_{th}$, where $m_{th}$ is the thermal mass. As described in
\cite{Garny:2009ni}, we also take the thermal non-Gaussian four-point correlation into account.
In addition, one momentum mode, taken to be $|\mathbf k|=3.0m_{th}$, is excited above the thermal occupation with an
effective particle number $\Delta n=7.5$ (see \cite{Lindner:2005kv}). For the numerical solution we use
a lattice of size $1400^2\times 32^3$ in $(x^0,y^0,\mathbf{k})$-space, a time-step $a_t=0.025/m_{th}$, and
momentum cut-off $\Lambda=\pi/a_x$ with $a_x=0.5/m_{th}$. Furthermore, we choose $\lambda_R/4!=0.75$
for the renormalized coupling and determine the counterterms according to \cite{Berges:2004hn}. The
numerical solutions conserve the total energy with an accuracy of better than $0.05\%$ (note that
this property would be lost when neglecting the back-reaction).

The plots show the dependence of the two-point functions on the relative time $\Delta t$ for the excited
momentum mode  $|\mathbf k|=3.0m_{th}$, as well as a mode with slightly smaller momentum $|\mathbf k|=2.96m_{th}$.
All correlators are evaluated at the fixed central time $t=17.5/m_{th}$. For the non-excited mode
shown in Figure~\ref{fig2}, we find that both the statistical propagator and the spectral function are
damped exponentially with respect to the relative time, as expected. Furthermore, the damping rate is
identical for both correlation functions. This property is expected from the fluctuation-dissipation relation
which is valid in thermal equilibrium. In addition, the relative
magnitude of the statistical propagator and the spectral function is close
to the expected value $1+2f_{BE}(\sqrt{{\mathbf k}^2+m_{th}^2}) \simeq 1.3$,
where $f_{BE}$ is the Bose-Einstein distribution function.

In Figure~\ref{fig1}, the corresponding propagators are shown for the excited momentum mode $|\mathbf{k}|=3.0m_{th}$.
The spectral function is also damped exponentially, whereas the statistical propagator shows a different
behaviour. However, the latter can be described to a good accuracy by the sum of the exponentially damped
equilibrium contribution and an undamped non-equilibrium contribution. This is precisely the behaviour
predicted by the analytical
solution~(\ref{distr:singular},\ref{distr:singular:coordinate}).
In addition, we have checked that when varying the central time $t$,
the non-equilibrium contribution to the statistical propagator decays exponentially, while the equilibrium
contribution remains constant. In addition, the decay rate of the non-equilibrium contribution obtained in
this way coincides with the damping rate of the equilibrium propagators. Again, this result is in accordance
with the analytical
solution~(\ref{distr:singular},\ref{distr:singular:coordinate}). We conclude that the analytical solution obtained under the thermal
bath assumption constitutes a reasonable approximation to the full numerical solutions, and is in particular
consistent with a non-equilibrium correlator that is undamped with respect to the relative time.

\section{Effectively Finite Width from the Gradient Expansion}
\label{section:effectivewidth}

We have now established analytically and numerically,
that when backreaction may be neglected,
the Wightman functions~(\ref{distr:singular}) that feature zero-width
out-of-equilibrium contributions are indeed correct solutions to the
Schwinger-Dyson equations on the CTP.
In this Section, we investigate,
whether the finite-width ansatz~(\ref{distr:smooth})
for ${\rm i}\Delta_\varphi^{<,>}$,
that we might intuitively expect and that it is often assumed
in the literature, may yet be appropriate, and if yes, in what sense.

For this purpose, we
substitute Eq.~(\ref{distr:singular}) into the right hand side of
Eq.~(\ref{KB:allorders}).
The substitution results in
\begin{align}
\label{Taylor:exp}
&\sum\limits_{n=0}^\infty\frac{1}{n!}
\left(-\frac{\rm i}2\Gamma_\varphi(\mathbf k)\right)^n
\left[(\partial_{k^0})^n \Pi_\varphi^{\cal A} \right]
\delta f_\varphi(k^0,\mathbf{k})
{\rm sign}(k^0)2\pi\delta(k^2-m_\varphi^2-\Pi_\varphi^H)
{\rm e}^{-\Gamma_\varphi(\mathbf k)t}
\,.
\end{align}
We now use the fact that we evaluate these terms in the
distributional sense, after integration over $d k^0$, such
that we can swap the $k^0$-derivatives through integration by
parts. Therefore, within the $\delta$-functions, we
replace $k^0\to k^0+\frac{\rm i}2 \Gamma_\varphi$, because
Eq.~(\ref{Taylor:exp}) corresponds to a Taylor series.
One may worry about the fact that we have deliberately replaced
expressions of $\Pi_\varphi^{\cal A}/k^0$ by $\Gamma_\varphi(\mathbf k)$ and therefore
truncated additional $k^0$ dependence. However, we may estimate derivatives
with respect to $k^0$ acting on these factors as yielding factors of $T^{-1}$,
such that the error incurred within each order $n$ of the gradient expansion
can be estimated as $(\Gamma_\varphi(\mathbf k)/T)^n$.
The same argument applies to derivatives acting on $\delta f_\varphi(k^0,k)$.
Derivatives acting on ${\rm sign} (k^0)$ are suppressed because
$\delta(k^2-m_\varphi^2-\Pi_\varphi^H)=0$ for $k^0=0$, at least provided
$m_\varphi^2+\Pi_\varphi^H\not=0$.
The suppression of gradients does
obviously not apply to derivatives acting on the $\delta$-function.
The consequences of this is what we explicitly calculate in the following.

From Eq.~(\ref{distr:singular:coordinate}), we see that
the $\delta$-function originates from the Fourier transform
of $\cos(\omega_\varphi(\mathbf k)\Delta t)$ with respect to $\Delta t$.
The real part of Eq.~(\ref{Taylor:exp})
enters the kinetic equations~(\ref{kinetic:allorders}),
while the imaginary part enters the
constraint equations~(\ref{constraint:allorders}).
To this end, $k^0$ has been assumed to be real, while now, we
aim to continue the expressions to complex $k^0$. This implies,
that we need to identify the various occurrences with $k^0$ or
${k^0}^*$. The correct prescriptions are determined by the causal
properties of the propagators, {\it i.e.} in two-time
representation, whether the time argument
is evaluated above or below the real axis, or in Wigner space,
where the poles are located with respect to the real axis of the
$k^0$-plane, {\it cf.} Appendix~\ref{appendix:poles} for a
quick reminder on this matter.
In two-time representation, the correct prescription
for the imaginary part
of the time argument is given by
\begin{align}
\label{continuation:propagator}
2 \Delta^{\cal A}_{\rm eff}(k)
=
&\frac{{\rm sign}(k^0)}{\omega_\varphi(\mathbf k)}\int\limits_{-\infty}^\infty d\,\Delta t\,
{\rm e}^{({\rm i}k^0-\frac12 \Gamma_\varphi(\mathbf k){\rm sign}(\Delta t))\Delta t}
\cos\left(
\omega_\varphi \Delta t
\right)
\\\notag
=&
\frac{{\rm sign}(k^0)}{2\omega_\varphi(\mathbf k)}
\sum\limits_\pm
\bigg[
\frac{\rm i}
{
k^0\pm\omega_\varphi(\mathbf k)
+\frac{\rm i}{2}\Gamma_\varphi(\mathbf k)
}
-
\frac{\rm i}
{
k^0\pm\omega_\varphi(\mathbf k)
-\frac{\rm i}{2}\Gamma_\varphi(\mathbf k)
}
\bigg]
\\\notag
=&
\frac{{\rm sign}(k^0)}{\omega_\varphi(\mathbf k)}
\frac{
\Gamma_\varphi(\mathbf k) ({k^0}^2+\omega_\varphi^2(\mathbf k))
}
{
({k^0}^2-\omega_\varphi^2(\mathbf k))^2+
[(k^0+\omega_\varphi(\mathbf k))^2+(k^0-\omega_\varphi^2(\mathbf k))^2]\frac14\Gamma_\varphi^2(\mathbf k)
+\frac1{16}\Gamma_\varphi^4(\mathbf k)
}
\\\notag
\approx&\frac{2\Gamma_\varphi(\mathbf k) k^0}
{({k^0}^2-\omega_\varphi^2(\mathbf k))^2+{k^0}^2\Gamma_\varphi^2(\mathbf k)}
\overset{\Gamma_\varphi\to 0}{=}
{\rm sign}(k^0)
2\pi\delta({k^0}^2-\omega_\varphi^2(\mathbf k))\,.
\end{align}
This result is purely real, such that it enters into
the kinetic equations, but not into the constraint equations.
Note that for these equations to be valid the actual integration interval
must be large compared to $\Gamma_\varphi^{-1}$. Regarding the replacement
${k^0}^2\to \omega^2_\varphi(\mathbf k)$ in the numerator, one may worry that
one drops a term that gives rise to relevant contributions to the integrations
over $k^0$. Note however, that within the collision term,
this finite width representation is
always
multiplied by distribution functions. We therefore assume, that all distribution
functions decay exponentially for large values of $|k^0|$, as it is the case for
{\it e.g.} the equilibrium distribution functions.

We also observe that
\begin{subequations}
\begin{align}
\Delta^R(k)&=\frac{1}{k^2-m^2+{\rm i}\varepsilon k^0}\,,
\\
\Delta^A(k)&=\left(\frac{1}{k^2-m^2+{\rm i}\varepsilon k^0}\right)^*\,,
\end{align}
\end{subequations}
may be continued to complex $k^0$ with ${\rm Im}[k^0]>0$ while
maintaining the causal properties, because these expressions
are analytic in the upper complex half-plane. In terms
of these propagators, we may express
\begin{align}
2\Delta^{\cal A}=
\frac{\rm i}
{k^2-m_\varphi^2-\Pi_\varphi^H(k)+{\rm i}\varepsilon k^0}
-\frac{\rm i}
{(k^2)^*-m_\varphi^2-\Pi_\varphi^H(k)-{\rm i}\varepsilon k^0}\,.
\end{align}
For real $k^0$ and infinitesimal $\varepsilon$, this corresponds
to the representation of
$2\pi\delta(k^2-m_\varphi^2-\Pi^H_\varphi(k))=2\Delta^{(0){\cal A}}$
as in Eq.~(\ref{Taylor:exp}). Through the Taylor expansion,
this expression can be directly continued to complex arguments
as $k^0 \to k^0+\frac{\rm i}{2}\Gamma_\varphi(\mathbf k)$
\begin{align}
\Delta^{\cal A}_{\rm eff}(k^0,\mathbf k)
=\Delta^{(0){\cal A}}
\left(k^0+\frac{\rm i}{2}\Gamma_\varphi(\mathbf k),\mathbf k\right)\,,
\end{align}
in agreement with Eq.~(\ref{continuation:propagator}).

It should be noted that due to the singular behaviour of the
on-shell
$\delta$-functions, it is not consistent to truncate the gradient
expansion of the collision term at any finite order in the diamond operator.
Here, we have shown that the gradients can easily be resummed to yield
a benign contribution with an intuitively clear meaning:
an effective finite-width distribution function. We therefore expect
that the systematics of the gradient expansion still applies to
kinetic theory, once singular terms arising from on-shell $\delta$-functions
are summed, following the methods introduced in the present Section.

In conclusion, the higher order contributions from the gradient expansion
that arise within the collision term precisely recover the na\"ive finite-width
ansatz~(\ref{distr:smooth})! In this sense, the fluctuation-dissipation
relation generalises to an out-of-equilibrium system of the scalar field
$\varphi$, for which backreaction can be neglected.
In turn, when we substitute
the finite width ansatz~(\ref{distr:smooth}) into the
collision term, we must not expand
to higher orders in gradients, as this is already accounted for
by the ansatz itself.

\section{Application to Processes with Kinematic Thres\-holds}
\label{section:thresholds}

Provided the self-energy $\Pi_\varphi(k)$ is not changing rapidly in $k^0$
close to
the pole, there is no leading order difference in whether we integrate
over the $\delta$-function or the finite width representation of the
propagator. This behaviour is however precisely what we assume within the
Breit-Wigner approximation~(\ref{Pi:Breit:Wigner}). Moreover,
when we integrate the kinetic equation~(\ref{kinetic:allorders})
for $k^0$ from $-\infty$ to $\infty$, both sides are trivially
zero, because of the symmetry properties of the real field.
Hence, in order to ``measure'' the effectively finite width
behaviour found in the previous Section, we yet need to specify
an interaction that can probe the off-shell contributions
of ${\rm i}\Delta^{<,>}_\varphi(k)$ for $k^0\not=\omega_\varphi(\mathbf k)$.

We therefore consider a reaction that
is mediated by a coupling $h$ and that is kinematically forbidden at tree-level.
For such a process, accounting for the effective finite width of the
propagators is indeed of leading order importance. Second, in order to
extract non-trivial information about the kinetic evolution, instead of plainly
integrating over $k^0$, one needs to include integrals over higher 
moments (additional powers of $k^0$ within the integrand).
For example, the distribution function
can be obtained from the integral
\begin{align}
f_\varphi(\mathbf k)=\int\frac{dk^0}{2\pi } |k^0|
{\rm i}\Delta_\varphi^<(k)-\frac12\,.
\end{align}
For different
observables, analogous prefactors for ${\rm i}\Delta_\varphi^<$
can be found that are typically
smooth functions of $k^0$ in the vicinity of the quasi-particle pole, such that
derivatives may be consistently neglected at leading order
in the gradient expansion.

For the purpose of probing the width of the field $\varphi$,
we introduce two additional scalar fields
$\chi_{1,2}$. The equation that determines the evolution of
$f_{\chi_1}$ follows from the kinetic equation~(\ref{kinetic:allorders})
(with the obvious replacement $\varphi\to \chi_1$) as
\begin{align}
\partial_t f_{\chi_1}(\mathbf k)
=&\int\frac{d k^0}{2\pi} |k^0| \partial_t{\rm i}\Delta^{<,>}_{\chi_1}
\\\notag
=&
-\int\frac{d k^0}{2\pi} {\rm sign}(k^0)
\frac12\cos(\diamond)
\left(
\{{\rm i}\Pi_{\chi_1}^>\}\{{\rm i}\Delta_{\chi_1}^<\}
-\{{\rm i}\Pi_{\chi_1}^<\}\{{\rm i}\Delta_{\chi_1}^>\}
\right)
\,.
\end{align}
Here, we have truncated terms that are of higher order in gradients
on the left-hand side.

In the presence of kinematic thresholds, including the
gradient terms that lead to the effective finite width can be
of leading importance.
In order to illustrate this, we consider an
interaction described by the Lagrangian term
$h \varphi \chi_1 \chi_2$, where $h$ is a real dimensionful coupling
constant. For the field $\chi_1$, this
induces the Wightman-type self-energies
\begin{align}
{\rm i}\Pi_{\chi_1}^{<,>}(k)=h^2\int\frac{d^4 p}{(2\pi)^4}
{\rm i}\Delta_\varphi^{<,>}(k-p)
{\rm i}\Delta_{\chi_2}^{<,>}(p)
\,.
\end{align}
Suppose now, that the field $\varphi$ deviates from equilibrium, as
described by the distribution function $\delta f_\varphi (\mathbf p)$,
while the fields $\chi_{1,2}$ are in equilibrium at the temperature $T$.
The fields have the masses $m_\varphi$ and $m_{\chi_{1,2}}$.
Moreover, assume that the production or decay rates of $\chi_{1,2}$ due
to the interaction with $\varphi$ are much smaller than the decay rate
$\Gamma_\varphi$. Then, we can use for ${\rm i}\Delta_{\chi_{1,2}}$
the on-shell propagators.
Moreover, the leading time dependence is in
$\delta f_\varphi (\mathbf p)$, and we may evaluate the 
action of the diamond operators as
\begin{align}
\partial_t f_{\chi_1}(\mathbf k)=&
-\frac12h^2{\rm Re}\bigg[\sum\limits_{n=0}^\infty
\int\frac{d^3 p}{(2\pi)^3}\int\frac{dk^0}{2\pi}\int\frac{dp^0}{2\pi}
\\\notag
&
{\rm i}\Delta_{\chi_2}^>(p)
\frac{1}{n!}
\left(\left[\frac{\rm i}2 \partial_t\right]^n
{\rm i}\Delta^>_\varphi(k-p)\right)
\partial_{k^0}^n{\rm i}\Delta_{\chi_1}^<(k)
{\rm sign}(k^0)
\bigg]
-(<,>\leftrightarrow >,<)\,.
\end{align}
We could now directly apply the Taylor expansion to the propagator
$\Delta_{\chi_1}$. However, since it is the field $\varphi$ which is
the origin of the
effective finite width behaviour, we rather perform an integration by parts,
in order to attach the derivatives $\partial_{k^0}$ to ${\rm i}\Delta_{\varphi}$.
We again ignore derivatives acting on $\Gamma_\varphi$, since these
would lead to terms that are suppressed by higher orders in the gradient expansion.
Note also that we can ignore the derivatives acting on the sign function,
since the on-shell propagator ${\rm i}\Delta_{\chi_1}(k)$ vanishes for
$k^0=0$.

Substituting the explicit forms of the propagator and performing the Taylor expansion
of ${\rm i}\Delta_{\varphi}(k-p)$ toward complex $k^0$, as it is explained in
the previous Section, we obtain
\begin{align}
\partial_t f_{\chi_1}(\mathbf k)
=&\frac12h^2\int\frac{d^3 p}{(2\pi)^3}\int\frac{dk^0}{2\pi}\int\frac{dp^0}{2\pi}
{\rm sign}(p^0)2\pi\delta(p^2-m_{\chi_2}^2)2\pi\delta(k^2-m_{\chi_1}^2)
\\\notag
\times&
\frac{
2\Gamma_\varphi(\mathbf k -\mathbf p) \omega_\varphi(\mathbf k -\mathbf p)
}
{
[(k-p)^2-m_\varphi^2]^2+(k^0-p^0)^2\Gamma_\varphi^2
}
\delta f_\varphi(\mathbf k-\mathbf p){\rm e}^{-\Gamma_\varphi(\mathbf k-\mathbf p)t}
\left[
\frac{1}{{\rm e}^{p^0/T}-1}
-\frac{1}{{\rm e}^{k^0/T}-1}
\right].
\end{align}
Now assume for definiteness, that $m_{\chi_1}+m_{\chi_2}>m_\varphi$ but
$m_{\chi_{1,2}}+m_\varphi>m_{\chi_{2,1}}$.
This is for example the case when all masses are of equal or similar size.
A $1\leftrightarrow 2$ process
between $\varphi$ and $\chi_{1,2}$ then is not possible provided all
particles are on-shell. In the present case, the field $\varphi$ obtains
an effective finite width, and $\partial_t f_{\chi_1}(\mathbf k)$ is non-zero,
even for the given mass relations. The above integral can in general
be evaluated numerically in
a straightforward way. Let us however additionally assume that
$m_{\chi_{1,2}},m_\varphi \gg T$ and 
$m_{\chi_1}+m_{\chi_2}-m_\varphi\ll m_{\chi_{1,2}},m_\varphi$. In that
situation, all particles are non-relativistic such that
$\omega_{\chi_{1,2},\varphi}(\mathbf k)\approx m_{\chi_{1,2},\varphi}$
and $\Gamma_\varphi(\mathbf k)\approx\Gamma_\varphi(\mathbf 0)$.
Moreover, the dominant reaction channel is
$\chi_1+\chi_2\leftrightarrow\varphi$, where $\varphi$ is off-shell.
The meaning of $\varphi$ being off shell is that it corresponds to a
virtual state, that eventually rescatters with addtional particles,
which is encompassed within the self-energy $\Pi^{\cal A}_\varphi$.
In this situation, the angular integration
can be performed easily.
Moreover, we note that in the integrand
${\rm sign}(p^0)\not={\rm sign}(k^0)$. With the above approximations, we
therefore obtain
\begin{align}
\partial_t f_{\chi_1}(\mathbf k)\!
\approx&\frac12h^2\!\frac{1}{2\omega_{\chi_1}(\mathbf k)}\!\!
\int\limits_0^\infty\!\frac{p^2dp}{(2\pi)^2}\frac{1}{2\omega_{\chi_2}(\mathbf p)}
\frac{
4\Gamma_\varphi(\mathbf 0)m_\varphi
}
{
\left[
m_{\chi_1}^2\!+m_{\chi_2}^2\!-m_\varphi^2
+2m_{\chi_1} m_{\chi_2}
\right]^2
\!+(m_{\chi_1}\!+m_{\chi_2})^2\Gamma_\varphi^2
}
\notag\\
\times&
\int\limits_{-1}^1 d\cos{\vartheta}\,
\delta f_{\varphi}(\mathbf k-\mathbf p){\rm e}^{-\Gamma_\varphi(\mathbf 0)t}\,,
\end{align}
where $\vartheta$ is the angle between $\mathbf k$ and $\mathbf p$
and where we have assumed that $\delta f_\varphi(\mathbf k)$ is
isotropic.
Clearly, this can be interpreted as a scattering process, 
between $\varphi$ and additional particles, that are described by
$\Gamma_\varphi$, via a virtual off-shell $\varphi$ into
$\chi_1$ and $\chi_2$.
In contrast, the result would be
strictly zero
if we would not have taken account of the resummation leading to the
effective
finite width propagator as discussed in the previous Section.
We emphasise that
for the present calculation, it is sufficient to take the Breit-Wigner
approximation for $\Pi^{\cal A}_\varphi(k)$, while the dependence of
$\Pi^{<,>}_{\chi_1}(k)$ on $k^0$ is of leading importance.
For related discussions of off-shell effects, {\it cf.}
Refs.~\cite{Garbrecht:2008cb,Drewes:2010pf}.

\section{Conclusions}
\label{section:conclusions}

For an out-of equilibrium scalar field, that interacts with a time-independent,
spatially homogeneous bath and that only experiences negligible backreaction,
we have reviewed and derived various aspects of the solutions to the
Schwinger-Dyson equations on the CTP. Our main focus is the solution for the
Wightman function~(\ref{distr:singular}) and its curious feature,
that it decomposes into a finite-width equilibrium and a zero-width
out-of-equilibrium contribution. The corresponding
Wightman function has been derived 
before in position space in Ref.~\cite{Anisimov:2008dz}, but no {\it direct}
interpretation of the different width of the two components is provided there.
The fact that the equilibrium component exhibits a finite width
is confirmed by a direct summation of the geometric series
of insertions of the self-energies into the propagators. This calculation
has first been performed in Ref~\cite{Altherr:1994jc},
and in Section~\ref{section:pinch}
we have reviewed it in order to put it within the present context.
The desirable cancellation of pinch singularities is accompanied
by a spectacular cancellation of the out-of-equilibrium contributions.
In order to verify that the vanishing width of the out-of equilibrium
component is not an artefact of the analytical approximations,
in Section~\ref{section:num},
we have presented a numerical study that clearly confirms
this feature. We emphasise that when using
the Schwinger-Dyson approach,
we do not encounter a problem with pinch singularities.
Besides, we do not
find a breakdown of perturbation theory applied to
out-of-equilibrium systems, that is sometimes associated with
pinch singularities~\cite{Greiner:1998ri,Boyanovsky:1999cy}.

The zero width behaviour of the out-of-equilibrium Wightman function,
that we have carefully confirmed calls for interpretation.
Intuitively, we expect that it should not be distinguishable,
whether a quasi-particle is part of the equilibrium or out-of-equilibrium
distribution. In Section~\ref{section:effectivewidth}, we present the
observation, that an effective finite width for the out-of-equilibrium
contribution
results from a summation of all gradients within the collision term.
Such a summation to all orders in gradients becomes necessary,
because derivatives acting on the on-shell $\delta$-function are
not suppressed and therefore do not comply with the usual suppression
of higher orders derivatives in the gradient expansion.
Section~\ref{section:thresholds}
provides an example, how the effective finite width is observable through
interactions, that are kinematically forbidden for zero-width particles.
Since the effective finite width takes the same form as the width of
the equilibrium propagator, our findings may be interpreted as a
generalisation of the fluctuation-dissipation relation to out-of-equilibrium
systems with negligible backreaction.

The results of the present work provide a justification for the use
of a heuristic finite-width ansatz, that is often intuitively applied
within kinetic theory, {\it e.g.} in
Refs.~\cite{Arnold:2000dr,Arnold:2001ba,Arnold:2001ms,Arnold:2002ja,Arnold:2002zm,Arnold:2003zc,Mrowczynski:1997hy,Garbrecht:2008cb}.
If one aims to derive kinetic theory from first principles of
Quantum Field Theory, using the CTP formalism, the present work fills in
a missing step in order to describe the finite width of out-of-equilibrium
particles. Note that the propagators for fermions reported in
Refs.~\cite{Anisimov:2010aq,Anisimov:2010dk} also feature a zero-width
non-equilibrium contribution. The
resummation technique leading to an effective finite width can in principle
be generalised to fermionic systems.

In the field of leptogenesis,
we see future applications in studying the role of the finite width
of the singlet neutrino, a discussion which has been initiated
in Refs.~\cite{Anisimov:2010aq,Anisimov:2010dk}. Moreover, the correct
description of the finite width of out-of-equilibrium neutrinos is
potentially crucial for
resonant leptogenesis~\cite{Flanz:1994yx,Flanz:1996fb,Pilaftsis:1997jf,Covi:1996wh,Pilaftsis:2003gt}
in a regime, where
the poles of the nearly mass-degenerate neutrinos overlap within
their width.

In conclusion, we have shown that the Schwinger-Dyson equations on the CTP
in the linear approximation, that allows for analytical solutions, are well
suited to describe finite-width effects for out-of-equilibrium systems.
Due to the zero-width behaviour of the Wightman functions, this is
not directly obvious, and to arrive at this result, a non-trivial summation
of gradients of all orders acting on on-shell $\delta$-functions
is necessary. Once this summation is performed, it appears that
the usual gradient expansion can be systematically extended order by order.
In the future, we may therefore expect further progress in the application
of the CTP formalism to the analytical description of out-of-equilibrium
systems.

\section*{Acknowledgements}

MG thanks Markus Michael M\"uller for his support. The work of MG was partially supported by the DFG
cluster of excellence ``Origin and Structure of the Universe.''
The work of BG is supported by the Gottfried Wilhelm Leibniz programme
of the Deutsche Forschungsgemeinschaft.

\noindent
{\bf Note added:} During the preparation of this Article,
Ref.~\cite{Fidler:2011yq}
appeared, which contains a similar argument about an effective
finite width, that does not arise from the relaxation toward equilibrium,
but from a time-dependent mass term.

\begin{appendix}

\section{Two Point Functions on the CTP and Wigner Space}
\label{appendix:twopoint}

Within the present work, we follow the notations and conventions
as given in Ref.~\cite{Prokopec:2003pj}. For a quick reference,
we quote here the relations between the two-point functions,
that are used within the present work, where $G$ may either stand for
a CTP Green function $\Delta$ or a  self-energy $\Pi$:
\begin{subequations}
\label{CTP:combinations}
\begin{align}
\label{CTP:advanced}
G^A=G^T-G^>=G^<-G^{\bar T}\quad&\textnormal{(advanced)}\,,
\\
\label{CTP:retarded}
G^R=G^T-G^<=G^>-G^{\bar T}\quad&\textnormal{(retarded)}\,,
\\
\label{CTP:hermitian}
G^H=\frac12(G^R+G^A)\quad&\textnormal{(Hermitian)}\,,
\\
\label{CTP:spectral}
G^{\cal A}=\frac1{2\rm i}(G^A-G^R)=\frac{\rm i}2(G^>-G^<)\quad&\textnormal{(anti-Hermitian, spectral)}\,.
\end{align}
\end{subequations}
The identification of these two-point functions with those
bearing CTP indices is
\begin{align}
G^T=G^{++}\,,\quad G^<=G^{+-}\,,\quad G^>=G^{-+}\,,\quad G^{\bar T}=G^{--}\,.
\end{align}
The functions $G^{<,>}$ are known as the Wightman functions,
$G^T$ is time ordered, $G^{\bar T}$ anti-time ordered.

The Wigner transform of a two-point function is defined as
\begin{align}
\label{Wigner:transform}
G(k,x)=\int d^4 r {\rm e}^{{\rm i}kr} G(x+r/2,x-r/2)\,,
\end{align}
where we refer to $r$ as the relative and to $x$ as the average coordinate.
In the situation of spatial homogeneity, there is no dependence on $\mathbf x$
(spatial translation invariance), such that $G(k,x)\equiv G(k,t)$.

\section{Prescription for the Imaginary Part of the Time Argument}
\label{appendix:poles}

The time-ordered and anti-time ordered Green functions
of a scalar field $\varphi$ are
\begin{align}
{\rm i}\Delta^{T,\bar T}(\Delta x)
=&\int \frac{d^4 p}{(2\pi)^4}{\rm e}^{-{\rm i}p\cdot\Delta x}
\frac{\rm i}{{p^0}^2-\omega^2_\varphi(\mathbf p)\pm{\rm i}\varepsilon}
\\\notag
=&\int\frac{d^3 p}{(2\pi)^3 2\omega_\varphi(\mathbf p)}
\bigg\{
\pm{\rm e}^{
{\rm i}(\omega_\varphi(\mathbf k)\mp{\rm i}\varepsilon)\Delta x^0
}\vartheta(\mp \Delta x^0)
\pm
{\rm e}^{
-{\rm i}(\omega_\varphi(\mathbf k)\mp{\rm i}\varepsilon)\Delta x^0
}\vartheta(\pm \Delta x^0)
\bigg\}
{\rm e}^{{\rm i}\mathbf p\cdot\Delta\mathbf x}
\,.
\end{align}
Thus, the prescription for the imaginary part always appears in such
a way that the integrand decays exponentially for large $|\Delta x^0|$.
Also recall that we may write
${\rm i}\Delta^{<,>}(\Delta x)
=\vartheta(\Delta x){\rm i}\Delta^{\bar T,T}(\Delta x)
+\vartheta(-\Delta x){\rm i}\Delta^{T,\bar T}(\Delta x)$.
This explains the explicit appearance of ${\rm sign}(\Delta t)$
within Eq.~(\ref{continuation:propagator}).

\end{appendix}

\end{document}